\documentclass[a4paper,man,natbib,floatsintext]{apa6}

\usepackage[english]{babel}
\usepackage[utf8x]{inputenc}
\usepackage{amsmath}
\usepackage{algorithm}
\usepackage[algorithmics]{}
\usepackage{apacite}
\usepackage{authblk}
\usepackage{float}
\usepackage{placeins}

\usepackage{algpseudocode}
\usepackage{graphicx}
\usepackage[colorinlistoftodos]{todonotes}
\algblockdefx[Foreach]{Foreach}{EndForeach}[1]{\textbf{foreach} #1 \textbf{do}}{\textbf{end foreach}}
\newcommand*{\BeginNoToc}{%
  \addtocontents{toc}{%
    \edef\protect\SavedTocDepth{\protect\the\protect\value{tocdepth}}%
  }%
  \addtocontents{toc}{%
    \protect\setcounter{tocdepth}{-10}%
  }%
}

\title{Wikibook-Bot - Automatic Generation of a Wikipedia Book}
\shorttitle{Wikibook-Bot - Automatic Generation of a Wikipedia Book}
\author{Shahar Admati, shahar.admati@gmail.com- corresponding author
 \\Lior Rokach liorrk@post.bgu.ac.il, 
 \\ Bracha Shapira, bshapira@bgu.ac.il}
 
 \affiliation{Ben-Gurion University of the Negev
 \\ P.O.B. 65 Beer-Sheva, Israel, 86501
 \\Tel. +97286477003, Fax +97286477527}

\abstract{A Wikipedia book (known as Wikibook) is a collection of Wikipedia articles on a particular theme that is organized as a book. 
\\We propose Wikibook-Bot, a machine-learning based technique for automatically generating high quality Wikibooks based on a concept provided by the user. In order to create the Wikibook we apply machine learning algorithms to the different steps of the proposed technique. Firs, we need to decide whether an article belongs to a specific Wikibook  - a classification task. Then, we need to divide the chosen articles into chapters - a clustering task - and finally, we deal with the ordering task which includes two subtasks: order articles within each chapter and order the chapters themselves. We propose a set of structural, text-based and unique Wikipedia features, and we show that by using these features, a machine learning classifier can successfully address the above challenges.
\\The predictive performance of the proposed method is evaluated by comparing the auto-generated books to existing 407 Wikibooks which were manually generated by humans. 
For all the tasks we were able to obtain high and statistically significant results when comparing the Wikibook-bot books to books that were manually generated by Wikipedia contributors.}

\begin{document}
 
\maketitle

\section{Introduction}
\begin{figure}[H]
  \centering
    \captionsetup{justification=}
    \includegraphics[width=0.3\textwidth, height=6cm]{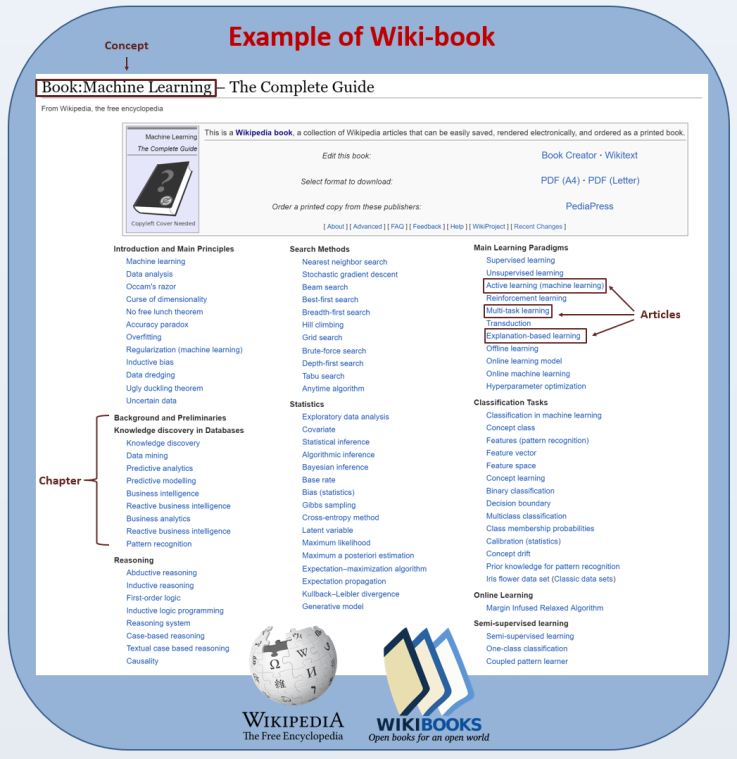}
    \caption{\footnotesize An example of an existing Wikibook -  "Machine Learning - The Complete Guide"\protect\footnotemark} \label{wikibook_example}
\end{figure}

Wikipedia is a free encyclopedia which is built collaboratively by contributors from all over the world. Wikipedia contains different types of pages: articles, category pages, books, etc. A Wikipedia book (known as a Wikibook) is a Wikipedia page that is based on a specific subject (concept) and contains references to various Wikipedia articles about the subject. The references are divided into chapters, which are rationally ordered (see example in figure \ref{wikibook_example}). 

A Wikibook can be consumed, free of charge, via a Wikipedia website. One can also purchase a hard copy of the book. These readily available and free resources enable online users to learn about a wide variety of subjects from different domains. Like all Wikipedia content, Wikibooks are created voluntarily by worldwide contributors who invest significant amounts of time and effort without recompense. The time-consuming process of writing a Wikibook is likely the main reason that as of early 2017, there were 5,325,457 articles in the English Wikipedia but only 48,775 Wikibooks. If the process of generating a Wikibook was automated, many more Wikibooks could be created, with little effort and in a short amount of time. Increasing the number of Wikibooks available for online consumption improves online user access to knowledge and information.

\footnotetext{https://en.wikipedia.org\/wiki\/Book:Machine\_Learning\_\%E2\%80\%93\_The\_Complete\_Guide}
We introduce Wikibook-Bot, which uses machine learning methods for the automatic generation of high quality Wikibooks based on a concept provided by the user. This tool represents the first attempt to fully automate the process of generating a book, without any human involvement. Attempts have been made to automate the generation of specific parts of a book, such as the index \cite{da2013integrating} and \cite{ait2006usingait2006using}. Other studies tried to develop an interactive system (e.g., BBookX \cite{pursel2016bbookx}) for generating a book in which the user chooses the content to include in it based on recommendations provided by the system. The novelty of our technique is that it is aimed at generating an entire Wikibook, without human involvement. 
Our technique applies machine learning algorithms from different families, such as classification (to decide whether an article belongs to a specific Wikibook or not), clustering (to divide the chosen articles into chapters), and ranking (order the chapters and articles); 

This paper describes the approach, which is evaluated by comparing the auto-generated books to existing Wikibooks which were created by humans. 

We demonstrate our technique for selecting the most relevant Wikipedia articles for a concept provided by a user. This task is challenging due of the sheer volume of articles that exist in Wikipedia and the need to select the most relevant articles among  millions of available articles. The selected articles will be part of the auto-generated Wikibook. We provide a detailed description of the candidate dataset creation and the feature engineering process. These features serve as input for the classification task aimed at deciding which articles to include in a specific Wikibook. We also present a method to divide the selected articles into chapters which will compose the auto-generated Wikibook.
We convert the clustering task into a classification task where we need to decide whether two articles 
should appear in the same chapter or not, utilizing  features extracted for each pair of articles. We also describe how the technique orders both the articles in each chapter and the chapters themselves.

The main contribution of this paper is our presentation and evaluation of a comprehensive approach for the automatic generation of Wikibooks, without any human involvement. In addition, we present and evaluate a set of structural, text-based and unique Wikipedia features, that enable accomplishment of the goals using machine learning techniques that we suggest for each of the involved steps. 

\section{Background and Related Work}
Attempts have been made to automate the generation of specific parts of a book, such as the index \citep{salton1988syntactic}, \citep{chang1992corpus}, \citep{da2013integrating} and \citep{ait2006using}. Other works describe efforts to automatically create a photo album given the user's selected photos \citep{wang2010automatic} and \citep{rabbath2011automatic}. BBookX \citep{liang2015bbookx}, \citep{liang2016bbookx} and \citep{pursel2016bbookx} is a system that has begun to automate the process of book generation, where the user chooses the content to include in a book based on recommendations provided by the system. The framework is based on a repository of books that are available on the Web, and Wikipedia articles that have been matched to chapters in printed books written by experts. In order to match the Wikipedia articles to the experts' book chapters, the system extracts the main concepts from the chapter's title using the Wikification method and ranks the results using an SVM machine learning algorithm. Using the BBookX GUI, in each step, the user provides a title and description that serves as the query for retrieving relevant chapters from the repository. Then, the user chooses a chapter from a list of retrieved chapters 
and has the opportunity to provide feedback which helps improve the query process. After submitting his/her choices, a book is created that includes the chapters selected by the user. While in this system, the user has a major role in the book generation process, 
our approach requires minimal human involvement only to provide the concept for the new Wikibook.

Apparently, the task of retrieving the relevant articles for a given subject can be formulated as a document clustering task. Prior research suggests various methods for document clustering \citep{jun2014document}, \citep{xie2013integrating} and \citep{jensi2014survey}. In these works, the starting point was the repository, which was divided into groups of documents that share a common implicit factor, and the research handled the challenge as an unsupervised problem. In contrast, we consider the problem of finding the most relevant articles for a given concept provided by the user as a supervised problem in which the concept provided by the user is the subject which serves as the unifying factor for all retrieved articles. 

Another study proposed Wiki-Write\citep{banerjee2016wikiwrite}, a system for the automatic generation of Wikipedia articles. Given a subject, Wiki-Write crawls the Web to find relevant unstructured content about it. In parallel, the system trains machine learning classifiers using content from existing Wikipedia articles on similar subjects in order to create a Wikipedia article; the system merges the Web-retrieved content for the new subject in relevant sections in a new Wikipedia article to produce a well-formed informative page.  
Wiki-write represents an attempt to automate the process of generating Wikipedia articles which can serve as components in Wikibooks.
\section{Generating the Wikibook Training Set}
To create the training and test sets required for utilizing our machine learning algorithms, we parsed the English Wikipedia dump from February 2016, saved the articles as text files, and extracted basic metadata such as article's categories, references, etc. There are approximately 6700 existing Wikibooks in the dump.
 
\begin{figure}[H] 
 \centering
    \captionsetup{justification=}
        \includegraphics[width=0.7\textwidth,height=6cm]{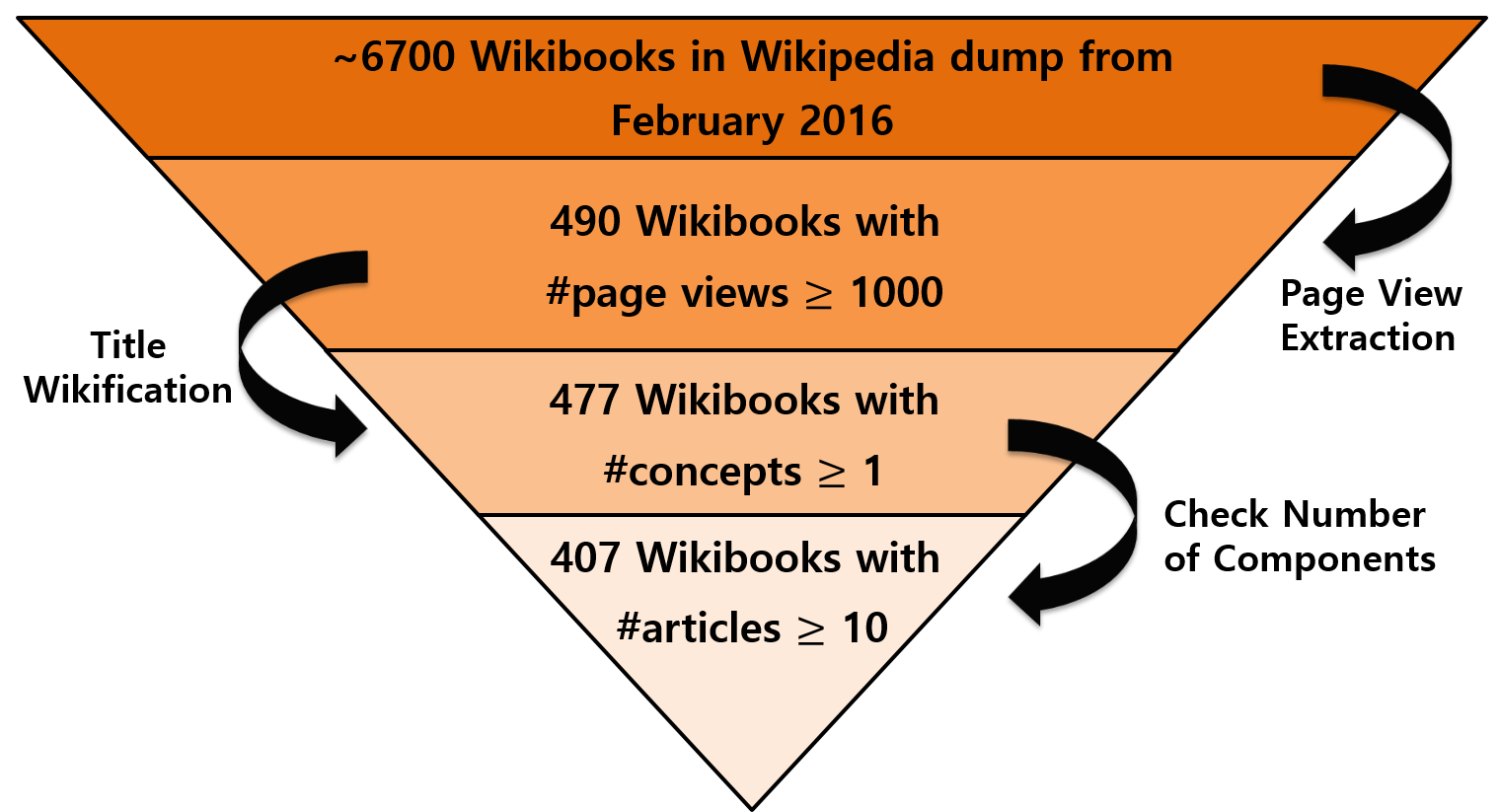}
    \caption{Filtering Wikibooks from English Wikipedia dump from February 2016} 
    \label{filtering}
\end{figure}

Since these existing Wikibooks form our gold standard both for training and testing our methods, we needed to ensure their  quality. We chose to concentrate on Wikibooks that were viewed at least 1000 times, (as inferred from the Wikipedia page views), during the six months period prior to the dump (8/1/15 - 2/1/16), based on the assumption that popular Wikibooks are of a reasonable quality. This filtering resulted in 490 Wikibooks. We applied Wikification on the titles of these Wikibooks using an existing API \citep{wikifier}. Wikification is the task of identifying concepts and disambiguating them into the corresponding Wikipedia pages \citep{cheng2013relational}. The Wikification ended successfully for 477 titles out of 490. We further filtered out Wikibooks with little content, i.e., Wikibooks with less than 10 components (articles); resulting with 407 wikibooks for training and testing our models (see figure  \ref{filtering}).

\section{The Proposed Approach}
\begin{figure}[H]
  \centering
      \captionsetup{justification=}
    \includegraphics[width=0.8\textwidth, height=13cm]{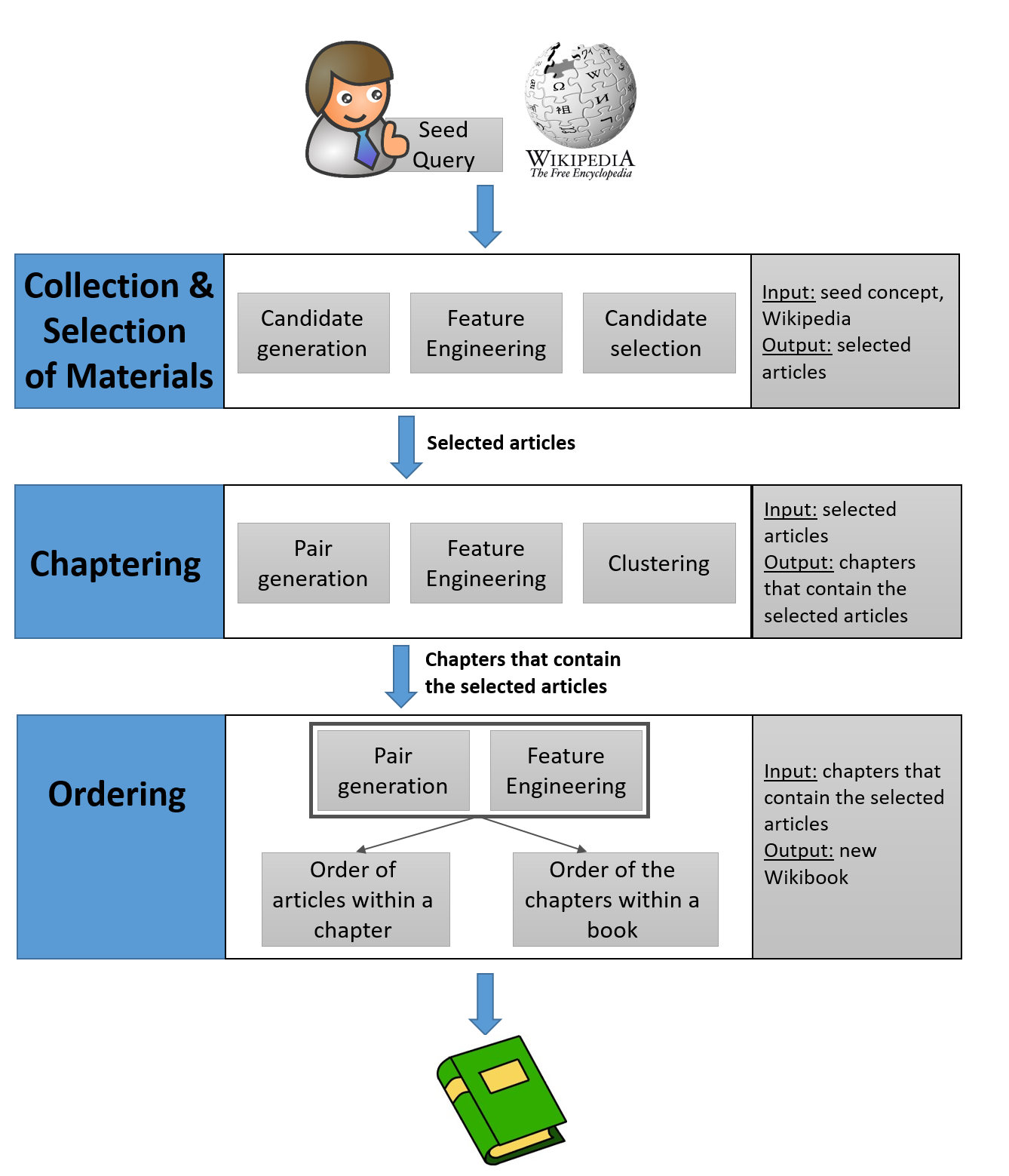}
    \vspace*{-8pt}
      \caption{\footnotesize Visual representation of the proposed approach} \label{flow}
\end{figure}

Our objective is to automatically generate high quality Wikibooks based on a user provided concept,  using machine learning methods. We present Wikibook-Bot, for automatic generation of Wikibooks. As described in 
figure \ref{flow} given a concept from the user, four main steps are performed: 1) seed query establishment, 2) collection and selection of materials,3) chaptering; and  4) ordering. Each step uses its own datasets which contain advanced features. 


For each step of the proposed technique, we present a comprehensive solution, along with an evaluation based on common machine learning and information retrieval measurements aiming at estimating the quality of the auto-generated Wikibook. 
Using the parsed dump, we created the necessary datasets for the application of the different steps: candidate datasets for the candidate selection task and pair datasets for both the chaptering and ordering tasks. 
%
\vspace*{-12pt}

\subsection{Step 1 - Establish Seed Query}
The first step 
is to obtain a seed query from the user that will serve as the concept of the book. This is the only point of interaction with the user. The seed query is the title for the book.We then use Wikification for extracting relevant seed concepts for the given seed query 
which results in a set of relevant Wikipedia articles and their references.

\subsection{Step 2 - Collection \& Selection of Materials}
\subsubsection{Candidate Generation\newline}\label{CandidateGeneration}
At this point all candidate articles that relate to the extracted seed concept must be retrieved. The candidate articles for each Wikibook, together with their characteristics, form the candidate dataset for this specific Wikibook. The candidate articles for the new auto-generated Wikibook are collected using the element of hops, which exist in Wikipedia if we consider it a network in which each article is a vertex, and hyperlinks between two articles are represented as edges. The inherent references to other Wikipedia articles that appear in the text of a Wikipedia article serve as an important part of its structure. 
If an article (i.e., article 1) is referred to by another article (i.e., article 2), article 1 is considered one hop away from article 2. In order to find candidate articles, we begin with the Wikipedia articles of the group of seed concepts, and identify articles which are at most three hops away from the seed concepts. Algorithm \ref{candidates} describes the process of identifying the candidate articles.
\begin{algorithm}[H]
\caption{Find candidates}\label{candidates}
\begin{algorithmic}[1]
\State $\textit{\small Apply Wikification on the user's seed query (the title of the Wikibook),}$
 \Statex $\textit{\small  in order to extract the main entities (i.e., seed concepts) from it.}$
\State $\textit{\small Identify the corresponding Wikipedia articles for the extracted seed concepts.}$
 \Statex $\textit{\small  These are the articles that we use to start collecting candidates from.}$
    \State $\textit{\small Select articles that are one, two or three hops away from the seed articles.}$
 \Statex $\textit{\small  Level 1 candidate articles are  articles referenced in the seed articles,}$
  \Statex $\textit{\small  level 2 candidate articles are articles referenced in level 1 articles}$
 \Statex $\textit{\small  level 3 candidate articles are articles referenced in the level 2 articles. }$
  \State $\textit{\small Combine the articles from levels 1, 2, and 3 as the candidate articles of the new Wikibook.}$
 
\end{algorithmic}
\end{algorithm}

After all candidate article datasets were created, we investigated whether the components of the original Wikibooks indeed appear in those datasets. We observed that more than 83\% of the datasets contain over 90\% of the components of the original Wikibook and 35\% of the datasets contain all of them. 


The size of the Wikibook candidate datasets ranges from 41,190 to 1,844,395 candidate articles for each Wikibook, while the number of components of the actual Wikibooks ranges from 10 to 570. Therefore, solving the selection of articles for a Wikibook from the available candidates as a classification task imposes a severe imbalance problem, where the minority class is the relevant articles, and the majority class consists of the non-relevant candidates. This fact affected the selection of algorithms to use for the various tasks.
\subsubsection{Feature Engineering\newline}\label{SelectionFeatureEngineering}
A wide variety of features are extracted for each candidate article in a dataset (see table \ref{features}). We extract three types of features: structural, text-based, and  unique Wikipedia features. Most of the features are computed relative to the seed concepts. Since the seed group may contain more than a single concept, each relative feature is represented by three different values (minimum, average, and maximum). In cases where the group of seed concepts includes only one concept, all three have the same value.

When extracting structural features, we consider Wikipedia a network. We construct a separate sub-network for each Wikibook from the candidate articles that were found to be connected with its seed concepts. We calculate structural measures \citep{fire2013computationally}, namely: in-degree, out-degree, pagerank, betweenness, closeness, hub, authority, and Dijkstra distance from the seed concept with the intuition that the connectivity between Wikipedia articles might reflect latent relation between them.

For the text-based features we computed features that represent the content of the Wikipedia articles as well as the structure of the text. We computed the cosine similarity between each candidate article and the seed concept, based on a document embedding model that is known to outperform bag-of-words models
\citep{le2014distributed}. We used a pre-trained Doc2Vec model \citep{lau2016empirical} which was trained over the English Wikipedia documents. The other text-based features consider the structure of the text. We calculated the relative difference and relative absolute difference between the seed concept and the candidate article for parameters, including the article's length and paragraph number. 

Each Wikipedia article has unique Wikipedia-based features such as categories, references, and page views. In addition to these basic unique features, we also calculated Pearson and Kendall Tau correlations between the seed concept's page view and the candidate article (both statistic and p-value), and the Jaccard coefficient on the categories of the seed concept and the candidate article. In addition, we computed the relative difference and relative absolute difference between the seed concept and the candidate article for parameters, including the number of categories and reference number.
\begin{table}[H]
\fontsize{8}{8}\selectfont
\begin{center}
\caption{Features of each candidate article in dataset}
\begin{tabular}{|p{3.5cm}|p{3.5cm}|p{3.5cm}|p{3.5cm}|}
\hline
In-degree & Out-degree & PageRank & Betweenness\\
\hline
Closeness & Hub & Authority & Min Dijkstra distance from the seed \\
\hline
Average Dijkstra distance from the seed concept & Max Dijkstra distance from the seed  & Min cosine similarity & Average cosine similarity\\
\hline
Max cosine similarity & Min length difference & Average length difference & Max length difference\\
\hline
Min absolute length difference & Average absolute length difference & Max absolute length difference & Min paragraph difference\\
\hline
Average paragraph difference & Max paragraph difference & Min absolute paragraph difference & Average absolute paragraph difference\\
\hline
Max absolute paragraph difference & Min Kendall Tau statistic & Average Kendall Tau statistic & Max Kendall Tau statistic\\
\hline
Min Kendall Tau p-value & Average Kendall Tau p-value & Max Kendall Tau p-value & Min Spearman statistic\\
\hline
Average Spearman statistic & Max Spearman statistic & Min Spearman p-value & Average Spearman p-value\\
\hline
Max Spearman p-value & Aggregated page views & Min Jaccard coefficient on categories & Average Jaccard coefficient on categories\\
\hline
Max Jaccard coefficient on categories & Min references difference & Average references  difference & Max references difference\\
\hline
Min absolute references difference & Average absolute references difference & Max absolute references difference & Min references to difference\\
\hline
Average references to difference & Max references to difference & Min absolute references to difference & Average absolute references to difference\\
\hline
Max absolute references to difference & Min categories number difference & Average categories number difference & Max categories number difference\\
\hline
Min absolute categories number difference & Average absolute categories number difference & Max absolute categories number difference & Classification\\
\hline
\end{tabular}
\label{features}
\end{center}
\end{table}

\subsubsection{Classification\newline}
In the classification task a decision is made whether to include each candidate article in the specific Wikibook (binary class). Each dataset corresponds to an existing Wikibook and contains the entire set of candidate articles for a specific Wikibook, along with its features. In this step, the class is added to each instance (candidate article):  1- if the article is part of the existing Wikibook its dataset represents; and 0, if otherwise. The datasets vary greatly with regard to the number of candidate articles, (41,190 to 1,844,395), as well as their size (in terms of memory), ranging from 28.5 MB to 1.19 GB; the number of components of the original Wikibooks ranges from 10 to 570. Therefore, due to the severe imbalance problem and the size of the datasets we use the LightGBM classifier\cite{LightGBM}. LightGBM is a gradient boosting framework that uses tree based learning algorithms that was designed with efficiency and distribution in mind and can handle large data efficienly. 
In the training phase, we built a separate model for each dataset. Then, we tested the models using the leave-one-out method. For each Wikibook (dataset), we predict which candidate articles will be included based on the other datasets' models 
Therefore, for each instance in the tested dataset we obtain $N-1$ predictions. LightGBM's outputs a probability that represents the confidence that the output class is 1 (i.e., included). See algorithm ~\ref{classification} below which outlines this process.
\begin{algorithm}[H]
\caption{Classification}\label{classification}
\begin{algorithmic}[1]
\Foreach{\small $D_i \in datasets$}
\State $\textit{\small Train and build LightGBM model}$
\EndForeach
\Foreach{\small $D_i \in datasets$}
\State $\textit{\small Predict the class of any instance in dataset $D_i$ based on all}$
  \Statex $\textit{\hspace{0.4cm} \small of the models except the model which was built based on $D_i$}$
  \Statex $\textit{\hspace{0.4cm} \small and get N-1 predictions as outputs}$
  \State $\textit{\small For each prediction output (out of N-1), sort its probabilities in}$
  \Statex $\textit{\hspace{0.4cm} \small descending order and rank so that the highest is ranked 1 and the }$
    \Statex $\textit{\hspace{0.4cm} \small lowest is ranked $n_i$ (number of instances in $D_i$)}$
    \State $\textit{\small Calculate the average ranking for each instance in $D_i$ based}$
  \Statex $\textit{\hspace{0.4cm} \small on all of the N-1 rankings}$
    \State $\textit{\small Build a logistic regression model with one feature: the}$
  \Statex $\textit{\hspace{0.4cm} \small average rankings}$
    \State $\textit{\small Predict the class of each instance in $D_i$}$
    \State $\textit{\small Output accuracy rates: AUC, precision, and recall}$
\EndForeach
\Foreach{\small $D_i \in datasets$}
    \State $\textit{\small Sort the instances in $D_i$ in ascending order according to the}$
      \Statex $\textit{\hspace{0.4cm} \small average ranking obtained in step 7}$
      \State $\textit{\small Take the first 20\% of the instances and create $D_i'$}$
\EndForeach
\Foreach{\small $D_i' \in datasets$}
    \State $\textit{\small Execute steps 5-10}$
\EndForeach
\end{algorithmic}
\end{algorithm}

\subsection{Step 3 - Chaptering}
After choosing the most relevant articles to be part of the new auto-generated Wikibook, the articles must be divided into groups that will form the new auto-generated Wikibook's chapters. For solving the chaptering task we concentrate on pairs of articles and examine their proximity. Based on the proximity we perform clustering which results in a division of articles into chapters.

\subsubsection{Pair Generation\newline}
For the chaptering task, we create new datasets for the 407 Wikibooks with features that are relevant for the current task; 
In order to enable "clean" evaluation of the chaptering task, we considered it  as a stand alone task (that can be integrated in the whole process of generating a Wikibook). For the sake of evaluation, the input of the chaptering task should not be affected by the results of the candidate selection task, but rather should receive perfect input. Thus, each of the 407 datasets that was generated for the chaptering task include only Wikipedia articles that are actually included in the original Wikibook, rather than candidates (regardless of whether they appeared as a candidate in the candidate dataset for this specific Wikibook or not).

The dataset for the chaptering phase consists of pairs of articles. Each instance in this type of dataset (referred to here as a pair dataset) represents a pair of articles, and their class represents whether these two articles appear in the same chapter in the Wikibook that the dataset represents or not. This instance definition allows us to transform the clustering task of dividing the articles into chapters into a classification task in which for each pair we have to decide whether the two articles in the pair should appear in the same chapter. In addition to the class, we extracted a wide variety of new features for each pair which describe the relationship between the two articles and may help to determine their class. 
Each pair dataset that relates to a Wikibook contains all possible combinations of two articles within the Wikibook; where the order in which the components are listed in the instance does not matter. Thus, each pair dataset contains ${N}\choose{2}$ instances, where $N$ is the number of components in the original Wikibook that the dataset represents.
 
\subsubsection{Feature Engineering\newline}\label{ChapteringFeatureEngineering}
For the pair datasets, the relationship between the two articles of each pair must be examined in order to determine if they should appear in the same chapter, in the specific Wikibook that this dataset represents. Therefore, in the pair datasets the features we extract (structural, text-based, and Wikipedia related features) for each pair of articles are relative to each other, in contrast to the candidate datasets where the features are relative to the seed concepts. For example, the difference between the number of paragraphs that exist in each of the articles in the pair rather than between a candidate article and its seed.

In the pair dataset, we use only the Dijkstra distance because it is the only relative feature in the structural features of the candidate datasets. In cases in which one (or even both) of the articles in a pair does not appear as candidates in the candidate dataset, and therefore are not part of the sub-network, the Dijkstra distance has a null value which is later replaced by the mean value of that feature in the specific pair dataset. 

In addition, we use all of the text-based features and the features that are unique to Wikipedia, except the aggregated amount of page views, which has an absolute value and is not a relative feature.Based on the relative features, we create the categorical features that are unique to the pair datasets. Each categorical feature represents a division that was made based on relative feature values. 
Each relative feature estimates the proximity between the articles in a different way,and based on this approximation we divide the articles into groups using three kinds of clustering algorithms: Diana (DIvisive ANAlysis Clustering) \citep{kaufman2009finding}, PAM (Partitioning Around Medoids) \citep{Reynolds2006} and Agnes (Agglomerative nesting) \citep{kaufman2009finding}. 

Based on each relative feature we create three categorical features, 
using each of the three different algorithms.  $N$ relative features result in $3N$ categorical features. The value of each of the three categorical features is determined based on whether the two articles in the pair appear in the same group according to the division that was made based on the approximation of proximity of the specific relative feature. If, the articles are in the same group, the value will be 1 and otherwise it will be 0. 
The process of creating the categorical features can be found in algorithm \ref{relative_feature_division}. In order to use these three clustering algorithms, the number of groups needs to be specified as input, 
we therefore used the number of chapters that exist in the original human-created Wikibooks. 
Nevertheless, because we aim to design an autonomous process, 
we also tried to automatically determine the number of groups using several criteria: Silhouette width \citep{rousseeuw1987silhouettes}, Calinski-Harabasz \citep{calinski1974dendrite}, Bayesian information criterion for expectation-maximization \citep{fraley2002model} and affinity propagation (AP) \citep{bodenhofer2011apcluster}. The AP algorithm outperformed all of the other algorithms, and its running time was much shorter than all the other algorithms. As expected, using the estimated number of groups instead of the actual number of chapters in the original Wikibook caused a moderate decline in the clustering accuracy measures.
The evaluation of the clustering task results using both the original and the estimated number of groups (i.e., chapters) can be found in section \ref{evaluation}.
\begin{algorithm}[H]
\caption{Create new categorical features based on the numeric relative features}\label{relative_feature_division}
\begin{algorithmic}[1]
\Foreach{\small $RF_i \in relative feature$}
\State $\textit{\small Normalize the proximity values to the range of 0-1:}$
  \Statex $\textit{\hspace{0.4cm} \small $nv=\frac{v-min_v}{max_v-min_v}$}$
\State $\textit{\small Transform the normalized proximity values into distances:}$
  \Statex $\textit{\hspace{0.4cm} \small $distance=1-nv$}$
\State $\textit{\small Build dissimilarity matrix based on the distances' values}$
\State $\textit{\small Divide the articles into groups using Diana algorithm}$
  \Statex $\textit{\hspace{0.4cm} \small based on the dissimilarity matrix}$
\State $\textit{\small Divide the articles into groups using PAM algorithm}$
  \Statex $\textit{\hspace{0.4cm} \small based on the dissimilarity matrix}$
\State $\textit{\small Divide the articles into groups using Agnes algorithm}$
  \Statex $\textit{\hspace{0.4cm} \small based on the dissimilarity matrix}$
  \Foreach{\small pair articles instance $P_j \in D_i$}
  \State $\textit{\small Set the value of the Diana relative feature division to}$
    \Statex $\textit{\hspace{0.8cm} \small 1 if the two articles appear in the same group based}$
    \Statex $\textit{\hspace{0.8cm} \small on the Diana division and 0 otherwise}$
      \State $\textit{\small Set the value of the PAM relative feature division to}$
    \Statex $\textit{\hspace{0.8cm} \small 1 if the two articles appear in the same group based}$
    \Statex $\textit{\hspace{0.8cm} \small on the PAM division and 0 otherwise}$
      \State $\textit{\small Set the value of the Agnes relative feature division to}$
    \Statex $\textit{\hspace{0.8cm} \small 1 if the two articles appear in the same group based}$
    \Statex $\textit{\hspace{0.8cm} \small on the Agnes division and 0 otherwise}$
  \EndForeach
\EndForeach
\end{algorithmic}
\end{algorithm}
\vspace*{-20pt}
\begin{algorithm}[H]
\caption{Clustering articles into chapters}\label{clustering}
\begin{algorithmic}[1]
\Foreach{\small $D_i \in datasets$}
\State $\textit{\small Train and build LightGBM model}$
\EndForeach
\Foreach{\small $D_i \in pair datasets$}
\State $\textit{\small Predict the class of any instance in dataset $D_i$ based on all}$
  \Statex $\textit{\hspace{0.35cm} \small of the models except the model which was built based on}$
  \Statex $\textit{\hspace{0.35cm} \small $D_i$ and get N-1 predictions as outputs}$
  \State $\textit{\small Calculate the average probability for each instance in $D_i$}$
    \Statex $\textit{\hspace{0.35cm} \small based on all of the N-1 predictions}$
\State $\textit{\small Convert the output average probabilities to distances:}$
  \Statex $\textit{\hspace{0.35cm} \small 1-average\_probability}$
\State $\textit{\small Build dissimilarity matrix based on the distances from step 7}$
\State $\textit{\small Use Agnes algorithm for dividing the articles in $D_i$ into chapters}$
  \Statex $\textit{\hspace{0.35cm} \small with the dissimilarity matrix from step 8 as an input}$
\State $\textit{\small Output accuracy rates: adjusted Rand index and p-value}$
\EndForeach
\end{algorithmic}
\end{algorithm}
\subsubsection{Clustering\newline}
In the clustering task we aim to divide the articles of each Wikibook into chapters. By creating the pair datasets we formalize the clustering as a classification task, in which each instance is a pair of articles and the class is based on whether the pair of articles should appear in the same chapter or not (binary). 
The pair datasets have similar characteristics to the candidate datasets 
so we can use the same classifier,
LightGBM. 
We train and build a model for each dataset and afterwards, we make predictions using the leave-one-out method. 
Predictions regarding which pairs of articles should appear together are made based on all of the other models except for the model of the tested dataset. 
The output of LightGBM classification is a probability  that the output class is 1 (i.e., the two articles in the pair should appear in the same chapter). Here, instead of converting the probabilities into ranks, we calculate the average probability for each instance based on the $N-1$ predictions we have, where $N$ is the number of datasets. Then, we transform the average probability into an average distance by calculating the complementary probability. Based on these distances, we build a dissimilarity matrix which serves as the input to the Agnes clustering algorithm.
Algorithm \ref{clustering} provides the detailed pseudo-code of the clustering process.
\subsection{Step 4 - Ordering}
The ordering task includes two subtasks: the ordering of the articles within each chapter, and the ordering between the chapters.
\subsubsection{Pair Generation\newline}
As done for the chaptering task, we develop and evaluate the solution for the ordering task as a stand alone module (that can be of course integrated in the whole process of generating a Wikibook). 
Therefore, for the sake of "clean" evaluation we assume perfect input which is not affected by the results of former phases, where all the articles that should be part of the Wikibook do exist in the given input and are already divided correctly into chapters.

We use the same pair datasets we created for the chaptering phase, with different features that fit the specific characteristics of the ordering task. The new generated pair datasets will be used for solving both, the within chapter and the between chapters ordering subtasks. 
In the ordering version of the pair dataset, each instance  represents a pair of articles, but the class of each instance represents whether the first article in the pair should appear before the second one in the Wikibook that the dataset represents, or not. This transforms the ordering task 
into a classification task in which for each pair a decision should be made regarding the correct ordering between the pair of articles. Based on the order defined by the article pairs we can derive the between chapters ordering, since the groups of articles compose the chapters.
\begin{table}[H]
\fontsize{8}{8}\selectfont
\begin{center}
\caption{\small Features of each pair of articles in dataset (ordering version)}
\begin{tabular}{|p{3.5cm}|p{3.5cm}|p{3.5cm}|p{3.5cm}|}
\hline
In-degree 1 & In-degree 2 & Out-degree 1 & Out-degree 2\\
\hline
PageRank 1 & PageRank 2 & Betweenness 1 & Betweenness 2\\
\hline
Closeness 1 & Closeness 2 & Hub 1 & Hub 2\\
\hline
Authority 1 & Authority 2 & Dijkstra distance between pair's articles & Cosine similarity between pair's articles\\
\hline
Length difference between pair's articles & Absolute length difference between pair's articles & Paragraph difference between pair's articles & Absolute paragraph difference between pair's articles\\
\hline
Kendall Tau statistic & Kendall Tau p-value & Spearman statistic & Spearman p-value\\
\hline
Aggregated page views 1 & Aggregated page views 2 & Jaccard coefficient on categories & References difference between pair's articles\\
\hline
Absolute references difference & References to difference between pair's articles & Absolute references to difference between pair's articles & Categories number difference between pair's articles\\
\hline
Absolute categories number difference between pair's articles & Classification & &\\
\hline
\end{tabular}
\label{orderingpairfeatures}
\end{center}
\end{table}
\subsubsection{Feature Engineering\newline}
We use all the features from the candidate datasets. If a feature is absolute, we transform it into two features,for each of the articles in the pair. For example, the structural feature in-degree will appear twice: in-degree 1, representing the in-degree of the first article in the pair and in-degree 2, for the second. If a feature is relative, its value is calculated to represent the relationship between the two articles.
The features are presented in table \ref{orderingpairfeatures}.
\begin{algorithm}
\caption{Ordering}\label{ordering}
\begin{algorithmic}[H]
\scriptsize
\Foreach{\footnotesize $D_i \in datasets$}
 \textit{\footnotesize Train and build LightGBM model} \EndForeach
\Foreach{\footnotesize $D_i \in datasets$} \Comment{Classification}
\State $\textit{\footnotesize Predict the class of any instance in dataset $D_i$ based on all of the models }$
    \Statex $\textit{\footnotesize except the model which was built based on $D_i$ and get N-1 predictions as outputs}$
  \State $\textit{\footnotesize For each prediction output (out of N-1), sort its probabilities in descending order and rank as follows:}$
  \Statex $\textit{\footnotesize the highest probability receives a rank of 1; the lowest is ranked $n_i$ (num. of instances in $D_i$) }$
    \State $\textit{\footnotesize Calculate the average ranking for each instance in $D_i$ based on all of the N-1 rankings}$
     \State $\textit{\footnotesize Build a logistic regression model with one feature: the average rankings}$
      \State $\textit{\footnotesize Predict the class of each instance in $D_i$}$
\EndForeach
\Foreach{\footnotesize $D_i \in datasets$} \Comment{Classification to ranking conversion}
\State $\textit{\footnotesize Set all articles' ranks to 0; Set all chapters' ranks to 0 }$
\Foreach{\small pair articles instance $P_j \in D_i$}
  \If {\footnotesize both articles in $P_j$ are in the same chapter} \Comment{Situation 1}
  \If {\footnotesize the predicted class for $P_j$ is 0}
      \State $\textit{\footnotesize increase the rank of the first article in $P_j$ by 1}$
  \Else
        \State $\textit{\footnotesize increase the rank of the second article in $P_j$ by 1}$
  \EndIf
  \Else \Comment{Situation 2}
    \If {\footnotesize the predicted class for $P_j$ is 0}
      \State $\textit{\footnotesize increase the rank of the chapter that contains the first article in $P_j$ by 1}$
         
  \Else
        \State $\textit{\footnotesize increase the rank of the chapter that contains the second article in $P_j$ by 1}$
            \EndIf
  \EndIf
\EndForeach
\State $\textit{\footnotesize Normalize chapters' ranks by dividing each chapter rank by the num. of chapter's articles}$
    \State $\textit{\footnotesize Sort the articles in ascending order according to their ranks}$
\State $\textit{\footnotesize Sort the chapters in ascending order according to their normalized ranks}$
    \State $\textit{\footnotesize Output accuracy rates for article ordering: Kendall Tau statistic and p-value}$
    \State $\textit{\footnotesize Output accuracy rates for chapter ordering: Kendall Tau statistic and p-value}$
\EndForeach
\end{algorithmic}
\end{algorithm}
\subsubsection{Order of Articles within a Chapter and between Chapters\newline}

Again , we use LightGBM. Here, we train and build a model for each dataset. Afterwards, we make predictions using the leave-one-out method,convert the LightGBM output probabilities into ranks and calculate average ranks based on all of the predictions. Based on the ranking we build a single logistic regression model for each dataset and make a single prediction for each instance (i.e., pair of articles) using the model. Next, we convert the binary classification into ranks. Based on the ranking we will sort the articles in each chapter and sort the chapters themselves. \looseness=-1

We distinguish between two types of ranks: the rank of an article in a chapter and the rank of a chapter within the book. Therefore, we also distinguish between two situations: 1) a situation in which the two articles in the pair appear in the same chapter; in this case the predicted class affects the rank of the articles
, and 2) a situation in which the articles appear in different chapters; in this case the predicted class affects the ranks of the chapters that include the pair of articles. 
If the articles appear in the same chapter 
the predicted class affects the order of articles within the chapter. In this situation, if the predicted class of a pair is 1,  the first article in the pair should appear before the second article in the pair, and therefore the rank of the second article should be increased by 1. If, 
the predicted class of a pair is 0,  the second article in the pair should appear before 
and  the rank of the first article should be increased by 1. The article order is dictated by an ascending order of the chapter's articles' ranks so the article within the chapter that obtained the highest rank will appear as the last article within the chapter.
If the articles appear in different chapters 
the predicted class affects the order of the chapters within a book. 
In this situation, if the predicted class of a pair is 1, it means that the chapter that contains the first article in the pair should appear before the chapter that contains the second article in the pair, and therefore the rank of the chapter that contains the second article should be increased by 1. 
If,  the predicted class of a pair is 0,  the chapter that contains the second article in the pair should appear before 
which implies that the rank of the chapter that contains the first article should be increased by 1. The chapter order is set by an ascending order of the chapter ranks.

For example, assume a Wikibook which contains two chapters: c1 which contains three articles a1, a2 and a3 and c2 which contains articles a4 and a5. Table \ref{examplepairdaraset} contains the pairs of articles, their chapters and the predicted class. Table \ref{chapterranks} and table \ref{articlesranks} contains the chapters' ranks and the articles' ranks respectively 
For example, article a1 is included in 4 pairs of articles that share the same chapter. However, only for one case the predicted class dictates that a1 should appear after another article (a3), resulting in rank 1 for article a1. 
 
There exist 6 instances for which the articles of the pair are not in the same chapter. In only one case, according to the predicted class, chapter c1 should appear after c2 and therefore its rank is 1. Chapter c1 includes 3 articles, therefore its normalized rank is $\frac{1}{3}$. According to the chapters' normalized ranks c1 will appear before c2 because it has a lower rank. Regarding the articles, in c1 the articles order will be a3, a1, a2 and in c2 the articles order will be a4 and then a5. 
\begin{table}[H]
\fontsize{8}{8}\selectfont
\begin{center}
\caption{Example of ordering prediction for pair dataset}
\begin{tabular}{|p{1.5cm}|p{1.5cm}|p{2cm}|p{2cm}|p{2.5cm}|p{3cm}|}
  \hline
  \textbf{article 1} & \textbf{article 2} & \textbf{chapter 1} & \textbf{chapter 2} & \textbf{predicted class} & \textbf{increment goes to}\\
  \hline
  a1 & a2 & c1 & c1 & 1 & a2\\
  \hline
  a1 & a3 & c1 & c1 & 0 & a1\\
  \hline
  a1 & a4 & c1 & c2 & 1 & c2\\
  \hline
  a1 & a5 & c1 & c2 & 1 & c2\\
  \hline
  a2 & a3 & c1 & c1 & 0 & a2\\
  \hline
  a2 & a4 & c1 & c2 & 1 & c2\\
  \hline
  a2 & a5 & c1 & c2 & 1 & c2\\
  \hline
  a3 & a4 & c1 & c2 & 0 & c1\\
  \hline
  a3 & a5 & c1 & c2 & 1 & c2\\
  \hline
  a4 & a5 & c2 & c2 & 1 & a5\\
  \hline
\end{tabular}
\label{examplepairdaraset}
\end{center}
\end{table}
\begin{table}[H] 
  \label{orderingexample} 
  \begin{minipage}[t]{0.5\linewidth}
  	\scriptsize
    \centering
    \caption{Chapters ranks}
    \begin{tabular}{|p{1.5cm}|p{2cm}|p{2.5cm}|}
    \hline
    \textbf{chapter} & \textbf{rank} & \textbf{normalized rank}\\
    \hline
    c1 & 1 & $\frac{1}{3}=0.3333$\\
    \hline
    c2 & 5 & 2.5\\
    \hline
    \end{tabular}

    \label{chapterranks}
    \vspace{4ex}
  \end{minipage}
  \begin{minipage}[t]{0.3\linewidth}
  	\scriptsize
    \centering
    \caption{Articles ranks}
    \begin{tabular}{|p{2cm}|p{2cm}|}
    \hline
    \textbf{article} & \textbf{rank}\\
    \hline
    a1 & 1\\
    \hline
    a2 & 2\\
    \hline
    a3 & 0\\
    \hline
    a4 & 0\\
    \hline
    a5 & 1\\
    \hline
    \end{tabular} 
    \label{articlesranks}
  \end{minipage} 
\end{table}

\section{Evaluation}\label{evaluation}

In order to test the proposed technique, we generated 407 groups of datasets corresponding to the 407 existing Wikibooks from the English Wikipedia dump from February 2016.

We evaluated the candidate selection process for each Wikibook using the AUC, precision, and recall measurements. In order to calculate  precision and recall , we examined the top $n$ articles according to the model where $n$ is the length of the original Wikibook. 
We examined these measures for each dataset separately and calculated the average AUC, average precision, and average recall based on all 407 datasets, and obtained the following results: AUC - 0.9765, precision - 0.2027, and recall - 0.2228. While the high AUC reflects the accuracy of the method, the precision and recall reflect the extreme imbalance that exists in the data. In one case, the model was required to select 92 relevant articles out of 1,844,394 candidates. If we randomly selected 92 articles out of 1,844,394, the default precision is $\frac{92}{1844394}=4.9881\cdot e^{-5}$ which is much lower than the value obtained by our method.

We also evaluated the chaptering process and determined whether the results are statistically significant using the adjusted Rand index and p-value measures. The p-value was calculated based on the alternative test (without simulations) 
as mentioned in \cite{qannari2014significance}. We use the notation of single asterisk (*) to indicate significance level of $\alpha<0.01$; double asterisk (**) when $\alpha<0.05$ and triple asterisk (***) when $\alpha<0.1$. We examined these measures for each dataset separately and calculated the average adjusted Rand index score and average p-value for all the 407 datasets together. When we used the number of chapters that exist in the original Wikibook as the number of groups parameter we obtained the following 
results: 1) using Agnes - adjusted Rand index score of 0.4276 with a p-value of 0.04939**, 2) using Diana - adjusted Rand index score of 0.389 with a p-value of 0.0478** and 3) Using PAM - adjusted Rand index score of 0.3233 with a p-value of 0.0714***. When using the estimated number of chapters calculated by the AP algorithm as the number of groups parameter we obtained slightly worse but still statistically significant, 
results: 1) using Agnes - adjusted Rand index score of 0.3716 with a p-value of 0.0303**, 2) using Diana - adjusted Rand index score of 0.3432 with a p-value of 0.0247** and 3) Using PAM - adjusted Rand index score of 0.3183 with a p-value of 0.0284**.
In both cases, Agnes algorithm outperformed all the other algorithms and therefore we chose it to be part of our proposed technique for solving the chaptering task.
We tried to estimate the number of chapters using different criteria: Silhouette width, Calinski-Harabasz, 
Bayesian information criterion for expectation-maximization and affinity propagation (AP), after some empirical trest and heuristics we we chose to use AP because it obtained the smallest heuristic value along with much shorter running time.

We evaluated the two ordering subtasks (ordering the chapters and the articles in each chapter) using the Kendall Tau and p-value measures. We examined these measures for each dataset separately and calculated the average Kendall Tau statistic and average p-value over the 407 datasets together. For the subtask of ordering the articles in each chapter, we obtained the following 
results: Kendall Tau statistic score of 0.8566 with a p-value of 0.0041.* For the subtask of ordering the chapters, we obtained the following 
results: Kendall Tau statistic score of 0.7735 with a p-value of 0.0686.*** The reason for the slightly worse significance level obtained for  ordering the chapters is that we believe that there could be more than one correct ordering of chapters.
For some books the order of many of the chapters does not matter, because there is not a direct link between the chapters. For example, in the Wikibook "Machine Learning - The Complete Guide" that was mentioned earlier (figure \ref{wikibook_example}), it does not matter if the clustering chapter appears before or after the classification chapter, because they are two separate  machine learning tasks. \looseness=-1


\section{Conclusions and Future Work}
we propose Wikibook-Bot, for automatically generating Wikibooks. We describe in detail each step of the process, which collects the candidate articles and accurately selects the most relevant articles for a given concept, among  millions of articles in the English Wikipedia; clusters the selected articles into chapters; and orders both the articles within each chapter and the chapters themselves. As can be observed from our results, each step of our method for generating automatic Wikibook is sound and provides meaningful outputs. Our proposed technique might be applicable for similar tasks such as automatic editing of printed books or conference proceedings. 
As future work we plan to apply this approach for creating new Wikibooks on subjects that do not exist in the English Wikipedia and publish them. Afterwards, we will monitor the page views and editions that may be made by Wikipedia contributors to the generated book for a long period of time. This will be a real world test for our approach. Future work may also include building a system based on our approach that can be used by Wikipedia contributors.

\bibliography{test}
\end{document}